\documentclass[aps,prb,reprint,amssymb,amsmath,superscriptaddress]{revtex4-1}

\usepackage[english]{babel}
\usepackage{graphicx}

\begin{document}

\title{Anisotropic conductivity and weak localization in HgTe quantum well with normal energy spectrum}

\author{G.~M.~Minkov}
\affiliation{Institute of Metal Physics RAS, 620990 Ekaterinburg,
Russia}

\affiliation{Institute of Natural Sciences, Ural Federal University,
620000 Ekaterinburg, Russia}

\author{A.~V.~Germanenko}

\author{O.~E.~Rut}
\affiliation{Institute of Natural Sciences, Ural Federal University,
620000 Ekaterinburg, Russia}

\author{A.~A.~Sherstobitov}
\affiliation{Institute of Metal Physics RAS, 620990 Ekaterinburg,
Russia}

\affiliation{Institute of Natural Sciences, Ural Federal University,
620000 Ekaterinburg, Russia}

\author{S.~A.~Dvoretski}

\affiliation{Institute of Semiconductor Physics RAS, 630090
Novosibirsk, Russia}

\author{N.~N.~Mikhailov}

\affiliation{Institute of Semiconductor Physics RAS, 630090
Novosibirsk, Russia}

\date{\today}

\begin{abstract}
The results of experimental study of interference induced
magnetoconductivity in narrow quantum well HgTe with the normal energy
spectrum are presented. Analysis is performed with taking into account
the conductivity anisotropy. It is shown that the fitting parameter
$\tau_\phi$ corresponding to the phase relaxation time increases in
magnitude with the increasing conductivity ($\sigma$) and decreasing
temperature following the $1/T$ law. Such a behavior is analogous to
that observed in usual two-dimensional systems with simple energy
spectrum and corresponds to the inelasticity of electron-electron
interaction as the main mechanism of the phase relaxation. However, it
drastically differs from that observed in the wide HgTe quantum wells
with the inverted spectrum, in which $\tau_\phi$ being obtained by the
same way is practically independent of $\sigma$. It is presumed that a
different structure of the electron multicomponent wave function for
the inverted and normal quantum wells could be reason for such a
discrepancy.
\end{abstract} 

\maketitle

\section{Introduction}
\label{sec:intr}

Two-dimensional systems based on gapless semiconductors HgTe are unique
object. HgTe is semiconductor with inverted ordering of $\Gamma_6$ and
$\Gamma_8$ bands. The $\Gamma_6$ band, which is the conduction band in
usual semiconductor, is located in HgTe lower in the energy than the
degenerate at $k=0$ band $\Gamma_8$, where $k$ is a quasimomentum. So
unusual positioning of the bands leads to crucial features of the
electron and hole spectrum under space
confinement.\cite{Volkov76,Dyakonov82,Lin85,Gerchikov90} For instance,
at some critical quantum well width, $d= d_c\simeq 6.5$~nm, the energy
spectrum is gapless and linear.\cite{Bernevig06}  In the wide quantum
wells, $d>d_c$, the lowest electron subband is mainly formed from the
$\Gamma_8$ states at small quasimomentum value, while the $\Gamma_6$
formes the hole states in the depth of the valence band. Analogously to
the bulk material such the band structure is referred to as inverted
structure. At $d<d_c$, the band ordering is normal. It is analogous to
that in conventional narrow-gap semiconductors; the highest valence
subband at $k=0$ is formed from the heavy hole $\Gamma_8$ states, while
the lowest electron subband is formed both from the $\Gamma_6$ states
and from the light  $\Gamma_8$ states.

The energy spectrum and specifics of transport phenomena in HgTe based
heterostructures were studied intensively last decade both
experimentally\cite{Landwehr00,Zhang02,Ortner02,Zhang04,Koenig07,Gusev11,Kvon11}
and theoretically.\cite{Bernevig06,Tkachov11,Nicklas11,Ostrovsky12}
Weak localization and specifics of the electron interference were
studied mainly theoretically for the range of parameters where the
energy spectrum is close to the Dirac type.\cite{Tkachov11,Ostrovsky12}
Analyzing symmetrical properties of the effective
Hamiltonian\cite{Bernevig06} and symmetrically relevant
symmetry-breaking perturbations the authors of
Ref.~\onlinecite{Ostrovsky12} show that the temperature and magnetic
field dependences of the interference induced magnetoresistance are of
a great variety. They can be localizing, antilocalizing or can
demonstrate the crossover from one type to another one depending on the
symmetry of perturbation and parameters of Hamiltonian.

Experimentally, the interference contribution to the conductivity in
two-dimensional HgTe heterostructures was studied in two papers only.
\cite{Olshanetsky10,Minkov12} Single quantum wells with 2D electron gas
were investigated in both papers. The authors of
Ref.~\onlinecite{Olshanetsky10} merely demonstrated that the
interference induced low-field magnetoresistivity is observed both in
narrow ($d<d_c$) and in wide ($d>d_c$) quantum wells. More detailed
studies were carried out in Ref.~\onlinecite{Minkov12}, however only
the structures with inverted spectrum, $d=(9-10)$~nm, were
investigated. There was found that the temperature dependences of the
fitting parameter $\tau_\phi$ corresponding to the phase relaxation
time show reasonable behavior, close to $1/T$. However, $\tau_\phi$
remains practically independent of the conductivity over the wide
conductivity range $(3-130)\,G_0$, where $G_0=e^2/\pi h$. This finding
is in conflict with theoretical arguments and experimental data for
conventional 2D systems with simple energy spectrum, in which
$\tau_\phi$ is enhanced with the conductivity. This fact calls into
question the adequacy of the use of the standard expressions for
description of the interference induced magnetoresistance in the
structures with inverted spectrum. Another possible reason for the
conflict is a specific of the phase relaxation in the such type of the
structures. The study of the interference induced magnetoresistance in
the HgTe quantum wells with normal spectrum ($d<d_c$) can shed some
light on this issue.

\section{Experimental details}
\label{sec:expdet}

\begin{figure}
\includegraphics[width=\linewidth,clip=true]{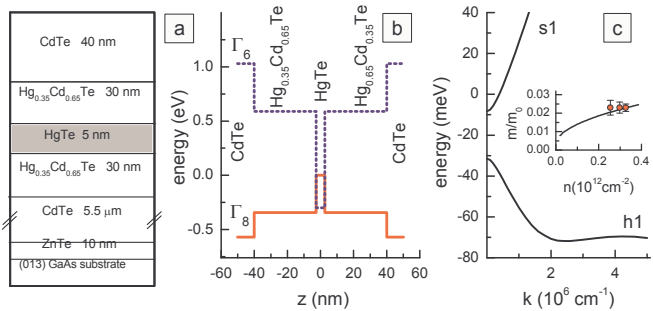}
\caption{(Color online) (Color online) Architecture (a) and energy  diagram (b)
of the structure under investigation. (c) The dispersion for the lowest electron
(s1) and highest hole (h1) subbands
calculated in the framework of isotropic $6\times6$ \textbf{kP} model. The inset shows
the electron density dependence of the effective mass for the electron subband s1. Symbols plot
the data, and the line shows the calculated dependence.}\label{f1}
\end{figure}

The HgTe quantum wells  were realized on the basis of
HgTe/Hg$_{1-x}$Cd$_{x}$Te ($x=0.65$) heterostructure grown by molecular
beam epitaxy on GaAs substrate with the (013) surface
orientation.\cite{Mikhailov06} The nominal width of the quantum well
was   $d=5$~nm. The architecture of the heterostructure, the energy
diagram, and dispersion law $E(k)$ calculated in the $6\times6$
\textbf{kP} model with the use of the direct integration technique as
described in Ref.~\onlinecite{Lar96} are shown in Fig.~\ref{f1}. The
parameters from Ref.~\onlinecite{Zhang01,Novik05} have been used. The
The samples were mesa etched into the Hall bars. The measurements were
performed on two types of the bars. The first type is the standard bar
shown in Fig.~\ref{f2}(a), the second one is the L-shaped two-arm Hall
bar schematically depicted in Fig.~\ref{f3}. To change  the electron
density ($n$) in the quantum well, the field-effect transistors were
fabricated with the parylene as an insulator and aluminium as a gate
electrode. Measurements were taken at the temperature of liquid helium.
All the data will be presented for $T=1.35$~K, unless otherwise
specified. The electron effective mass needed for the quantitative
interpretation was determined from the analysis of the temperature
behavior of the Shubnikov-de Haas (SdH) oscillations. It is
approximately equal to $0.023\,m_0$ within the range $n=(2.5-3.3)\times
10^{11}$~cm$^{-2}$. These data agree well with theoretical results
obtained  in the \textbf{kP} model [see inset in Fig.~\ref{f1}(c)]. For
lower electron density, $n<2.5\times 10^{11}$~cm$^{-2}$, we employ the
theoretical $m$~vs~$n$ dependence.

\section{Results and discussion}
\label{sec:res}

\begin{figure}
\includegraphics[width=\linewidth,clip=true]{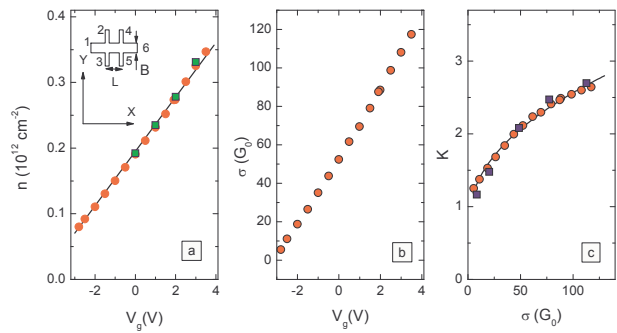}
\caption{(Color online) The gate voltage dependences of (a) electron
density and (b) conductivity obtained from the measurements on the standard Hall bar,
shown in the inset. Circles and squares in panel (a) are data
obtained from the Hall and SdH effect, respectively. (c) The conductivity dependences
of the conductivity anisotropy obtained from the measurements of nonlocal conductivity
(squares) and from the measurements performed on the L-shaped Hall bar (circles).
The lines are provided as a guide to the eye.}\label{f2}
\end{figure}

The gate voltage ($V_g$) dependence of the electron density (found both
from the Hall effect and from the SdH oscillations) and conductivity
are plotted in Figs.~\ref{f2}(a) and \ref{f2}(b). One can see that $n$
linearly changes with $V_g$  with the  rate $dn/dV_g$ of about
$(4.1\pm0.1)\times 10^{10}$~cm$^{-2}$V$^{-1}$. This rate is close to
$C/e$, where $C$ is specific capacitance measured between the gate
electrode and 2D gas for the same structure. As seen we were able to
change the conductivity  from $5\,G_0$ to $120\,G_0$ in  our $V_g$
range.

Because the heterostructures were grown on GaAs substrate with (013)
surface orientation, it is naturally to suppose that walls of the HgTe
quantum well are not ideal plane and can be corrugated. For thin
quantum wells (for our case the nominal quantum well width consists of
only of  $7-8$ lattice constants)  such corrugation can result in the
anisotropy of conductivity.\footnote{It should be mentioned that not
only the corrugation of the HgTe/Hg$_{1-x}$Cd$_x$Te interfaces may
result in the conductivity anisotropy. Anisotropy can be of
technological origin. For instance, the quasiperiodic structure of
different composition of solid solution Hg$_{1-x}$Cd$_x$Te can be
formed during epitaxial growth under certain conditions as shown in
Ref.~\onlinecite{Sabinina11}. This modulated composition can also
result in anisotropy of electrical properties.  } Therefore, let us
consider the results concerning the conductivity anisotropy before to
analyze the interference induced magnetoresistivity quantitatively.
They have been obtained by two methods.

The first method is based on  measurements of the nonlocal conductance
on the standard Hall bar. When the principal axes $x$ and $y$ of the
conductivity tensor coincide with axes $X$ and $Y$ of the sample
coordinate system [see the inset in Fig.~\ref{f2}(a)] the conductivity
anisotropy $K=\sigma_{xx}/\sigma_{yy}$ can be found from the ratio of
nonlocal conductance ($G_{nL}$) to the local one
($G_{L}$):\cite{German90}
\begin{equation}
 \frac{G_{nL}}{G_{L}}=\frac{4\sqrt{K}}{\pi} \exp\left(-\frac{L}{B}\frac{\pi}{\sqrt{K}} \right),
 \label{eq01}
\end{equation}
where $B$ and $L$ stand for the width of the Hall bar and for the
distance between probes $3$ and $5$, respectively, $G_L= I_{16}
/V_{35}$ ($I_{16} /V_{24}$), $G_{nL}= I_{23}/V_{45}$ with $I_{ik}$ as
the current flowing through the probes $i$ and $k$, and $V_{lm}$ as the
voltage drop between the probes $l$ and $m$.

The conductivity dependence of $K$ measured by this method is plotted
by squares in Fig.~\ref{f2}(c). It is seen that the conductivity
anisotropy increases with the increase of the conductivity and electron
density, and it reaches the value of $K\simeq 2.7$ at $\sigma=120\,G_0$
($n=3.5\times10^{11}$~cm$^{-2}$). Of course, such a dependence may
result from an extended mechanical defects, such as scratches or
notches, directed along the bar. To make sure that it is not the case
and the anisotropy of the conductivity is the physical property of 2D
electron gas in the structures investigated,  we used the second
method.

\begin{figure}
\includegraphics[width=0.9\linewidth,clip=true]{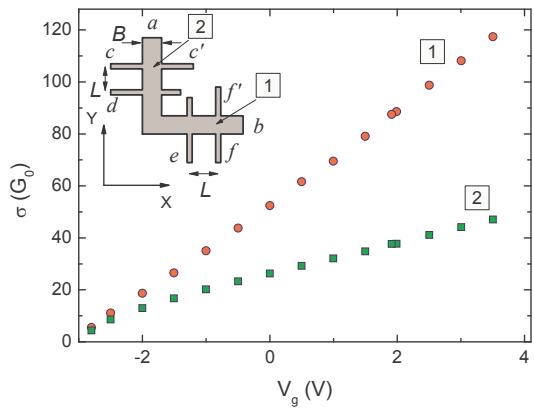}
\caption{The gate voltage dependence of the conductivity measured on the
different arms of the L-shaped Hall bar shown in the inset.}\label{f3}
\end{figure}

The anisotropy value in the second method is obtained from the
measurements performed on the L-shaped Hall bar (see the inset in
Fig.~\ref{f3}), which is made on the basis of the same wafer in such a
way that the orientation of arm $1$ on the wafer coincides with the
orientation of the Hall bar shown in the inset in Fig.~\ref{f2}(a).
Such measurements show that the electron densities in both arms are
equal to each other with the accuracy better than $3$\%, but the
conductivity of arm $1$ ($\sigma_1=I_{ab}/V_{ef}\times
L/B=\sigma_{xx}$) is significantly higher than that of arm $2$
($\sigma_2=I_{ab}/V_{cd}\times L/B=\sigma_{yy}$) as shown in
Fig.~\ref{f3}. The ratio $\sigma_1/\sigma_2=\sigma_{xx}/\sigma_{yy}=K$
plotted  against the conductivity of arm $1$ in Fig.~\ref{f2}(c) shows
that the $K$ values obtained by this method practically coincide with
those obtained by the first method.

Thus, the conductivity of electron gas in the studied heterostructures
with narrow HgTe quantum well grown on (013) substrate is strongly
anisotropic that should be taken into account at analysis of  transport
properties of such type structures.

\begin{figure}
\includegraphics[width=\linewidth,clip=true]{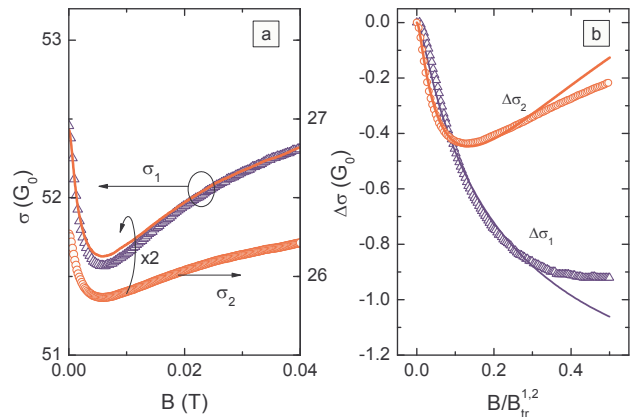}
\caption{The dependences of (a) $\sigma_{1,2}(B)$ and (b) $\Delta\sigma_{1,2}(B/B_{tr}^{1,2})$
for the L-shaped Hall bar measured for $V_g=0$~V at $T=1.4$~K. $B_{tr}^{1}=0.014$~T, $B_{tr}^{2}=0.056$~T.
The curve in the panel (a) is $\sigma_2(B)\times K=\sigma_2(B)\sigma_1(0)/\sigma_2(0)=2\,\sigma_2(B)$.
The curves in the panel (b) are the results of the best fit by Eq.~(\ref{eq05}) with the parameters given
in the text.}\label{f4}
\end{figure}

We are now in a position to consider the low field magnetoconductivity.
The magnetic field dependence of $\sigma_{1}$ and $\sigma_{2}$ measured
for both arms of L-shaped Hall bar at $T=1.4$~K and $V_g=0$~V are
presented in Fig.~\ref{f4}(a). Qualitatively, these dependences are
analogous. The conductivity decreases in the low magnetic field,
reaches the minimum near $B\simeq 60$~mT and increases at higher
magnetic field. Such a behavior is typical for the interference induced
magnetoconductivity for the case of  fast spin relaxation,
$\tau_s<\tau_\phi$, where $\tau_s$ is the spin relaxation time. Notice
that the magnitude of the magnetoconductivity
$\Delta\sigma_{1}(B)=\sigma_{1}(B)-\sigma_{1}(0)$ and
$\Delta\sigma_{2}(B)=\sigma_{2}(B)-\sigma_{2}(0)$ is different whereas
the characteristic magnetic field scales are the same; $\sigma_{2}$
being multiplied by the factor close to two  coincides practically with
$\sigma_{1}$ as illustrated by solid line in Fig.~\ref{f4}(a).

Theoretically, the weak localization and antilocalization for narrow
gap ($d\simeq d_c$) HgTe quantum wells is comprehensively studied in
Ref.~\onlinecite{Ostrovsky12}. The effective
Hamiltonian\cite{Bernevig06} is used to describe the spectrum of the 2D
gas. The authors show that the magnetoconductivity can be positive or
negative or it can demonstrate alternative-sign behavior with the
increasing magnetic field. Which scenario is realized, it depends on
the  position of the Fermi level as illustrated in
Ref.~\onlinecite{Ostrovsky12} by Fig.~1. For our samples, the Fermi
level lies in the linear part of the spectrum. The Fermi energy equal
to  ($25-55$)~meV for different $V_g$ is always larger than the value
of $m(k_F)\simeq (15-20)$~meV [$m(k_F)$ is introduced by Eq.~(4) in
Ref.~\onlinecite{Ostrovsky12}]. For such the case the 1Sp$\to$2U
scenario  should be realized with the growing magnetic field according
to Ref.~\onlinecite{Ostrovsky12}. This means that the
magnetoconductivity being negative in the low field has to saturate in
the higher one. As seen from Fig.~\ref{f4} the experimental picture of
the magnetoconductivity is somewhat different. The magnetoconductivity
is really negative in the low magnetic field, $B\lesssim 60$~mT.
However, in the higher magnetic field the rise of the conductivity is
observed instead of the saturation. In Ref.~\onlinecite{Ostrovsky12},
such a behavior (characteristic to the pattern 1Sp$\to$2O) was
predicted only for very low and very large concentrations of carriers,
corresponding to the strong nonlinearity of the spectrum.  The
difference apparently results from the fact that only the diffusion
(logarithmic) contributions to the interference quantum correction have
been taken into account in Ref.~\onlinecite{Ostrovsky12}. As seen from
Fig.~\ref{f4} the rise of the conductivity is evident in the magnetic
field which is not very small as compared with the characteristic for
weak localization transport magnetic field, $B_{tr}=\hbar/4eD\tau$,
where $\tau$ is the transport relaxation time and $D$ is the diffusion
constant. This indicates that the ballistic contribution which also
depend on magnetic field may play significant role in the actual case.
Indeed, when the leading logarithmic contributions cancel each other
(regime 2U), the subleading ballistic terms may dominate the
magnetic-field dependence of the interference correction to the
conductivity. In fact, the magnitude of the negative magnetoresistance
in Fig.~\ref{f4} does not exceed $G_0$ and can therefore result from
the ballistic effects.

Thus, unfortunately the expressions from Ref. 15 cannot be applied to
analyze our data quantitatively in the whole range of magnetic fields.
There is a clear need of a theoretical description of
magnetoconductivity in HgTe structures in a whole range of magnetic
fields including ballistic effects. Therefore, we will follow
traditional way using the Hikami-Larkin-Nagaoka (HLN)
expression\cite{Hik80,Knap96} to describe the interference induced
magnetoconductivity:
\begin{equation}
\Delta\sigma(B)= G_0 {\cal H}\left(\frac{B}{B_{tr}},\frac{\tau}{\tau_\phi},\frac{\tau}{\tau_s}\right), \label{eq05}
\end{equation}
where
\begin{eqnarray}
{\cal H}(b,x,y) &=&\psi\left(\frac{1}{2}+\frac{x+y}{b}\right)-\ln{\left(x+y\right)} \nonumber\\
&+&\frac{1}{2}\psi\left(\frac{1}{2}+\frac{x+2y}{b}\right)-\frac{1}{
2}\ln{\left(x+2y\right)}\nonumber\\
&-& \frac{1}{2}\psi\left(\frac{1}{2}+\frac{x}{
b}\right) +\frac{1}{2}\ln{\left(x\right)}-\psi\left(\frac{1}{2}+\frac{1}{b}\right) \nonumber
 \label{eq10}
\end{eqnarray}
with $\psi(x)$ as a digamma function.

Let us firstly analyze the data obtained on the two arms  by the
standard manner as if they have been obtained on the two different
samples with isotropic conductivity. The results of the best fit made
within the magnetic field range $B/B_{tr}^{1,2}=0-0.3$
($B_{tr}^{1}=0.014$~T, $B_{tr}^{2}=0.056$~T) for each arm are presented
in Fig.~\ref{f4}(b). The figure shows a good fit of the equation to the
data. Nevertheless, the values of the fitting parameters are different.
While the difference between the $\tau_\phi$ values for the arms 1 and
2 is not very large ($\tau_\phi^1=3.8\times 10^{-11}$~s and
$\tau_\phi^2=3.2\times 10^{-11}$~s), the difference between $\tau_s^1$
and $\tau_s^2$ is significant: $\tau_s^1=0.9\times 10^{-12}$~s is
approximately  five times smaller than $\tau_s^2=4.7\times 10^{-12}$~s.
In what follows we show that the reason for such a discrepancy is
neglect of the  conductivity anisotropy in the above data analysis.

The interference correction to the conductivity of the 2D anisotropic
systems was studied in Refs.~\onlinecite{Woelfle84,Bhatt85}. If one
follows this line of attack, the interference induced
magnetoconductivity can be written within the diffusion approximation,
$\tau_\phi$, $\tau_s\gg \tau$, in the form:
\begin{eqnarray}
\Delta\sigma_{xx}(B)&=&\sqrt{K}\,G_0\, {\cal H}\left(\frac{B}{B'_{tr}},\frac{\tau'}{\tau_\phi},\frac{\tau'}{\tau_s}\right)\nonumber\\
\Delta\sigma_{yy}(B)&=&\frac{1}{\sqrt{K}}\,G_0\, {\cal H}\left(\frac{B}{B'_{tr}},\frac{\tau'}{\tau_\phi},\frac{\tau'}{\tau_s}\right),
 \label{eq20}
\end{eqnarray}
where $B'_{tr}=\sqrt{B_{tr}^{1}B_{tr}^{2}}$ and
$\tau'=\sqrt{\tau_1\tau_2}$.

\begin{figure}
\includegraphics[width=0.7\linewidth,clip=true]{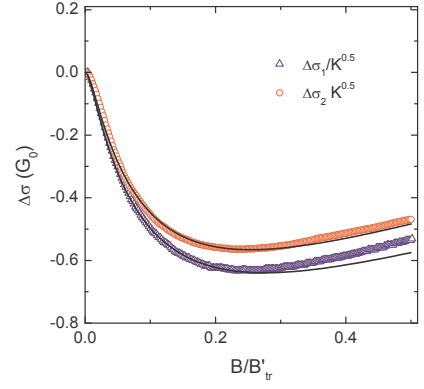}
\caption{The values of $\Delta\sigma_1/\sqrt{K}$ and
$\Delta\sigma_2\sqrt{K}$ plotted as  functions of
$B/B'_{tr}$, $B'_{tr}=0.028$~T, for $V_g=0$~V and $T=1.4$~K. The symbols are the data,
the solid lines are the result of the best fit by Eq.~(\ref{eq05}) with the parameters given in the text.}\label{f5}
\end{figure}

Thus,  the values of $\Delta\sigma_1/\sqrt{K}$ and
$\Delta\sigma_2\sqrt{K}$ plotted as  functions of $B/B'_{tr}$ should
coincide and they should be described by the expression,
Eq.~(\ref{eq05}), with $b=B/B'_{tr}$, $x=\tau'/\tau_\phi$, and
$y=\tau'/\tau_s$. As evident from Fig.~\ref{f5} the experimental data
for the arms 1 and 2 replotted in such the manner are really close to
each other, but, what is more important, we obtain the very close
parameters for two arms: $\tau_\phi=3.8\times 10^{-11}$~s, $3.5\times
10^{-11}$~s and $\tau_s=3.1\times 10^{-12}$~s, $3.2\times 10^{-12}$~s
for arms $1$ and $2$, respectively. Existing difference between the
data for different arms and between the corresponding fitting
parameters may result from the fact that the diffusion regime,
$\tau\ll\tau_s,\,\tau_\phi$,  is not strictly realized under our
experimental conditions; $\tau_s$ is only seventeen times larger than
$\tau'$.

It is pertinent to note here that the $\tau_\phi$ values are very close
to each other independently of that how they have been obtained.
Considering the arms 1 and 2 as independent isotropic samples and
taking into account the conductivity anisotropy we have obtained
practically the same results: $\tau_\phi=(3.2-3.8)\times 10^{-11}$~s.
As for the spin relaxation time, the ignoring of the conductivity
anisotropy when treating the data can give an error in $\tau_s$ of
several times of magnitude.

Now, making sure that  Eq.~(\ref{eq20}) for the interference correction
in anisotropic 2D system  describes the experimental data adequately we
can proceed to analysis of the dependences of the parameters
$\tau_\phi$ and $\tau_s$ on the temperature and conductivity.

\begin{figure}
\includegraphics[width=0.7\linewidth,clip=true]{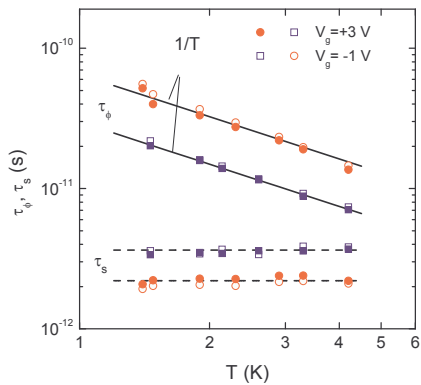}
\caption{The temperature dependences of $\tau_\phi$ and $\tau_s$ for two gate voltages, $V_g=-1$~V and $3$~V.
Solid and open symbols plot the data obtained from analysis of the dependences $\Delta\sigma_1(B)/\sqrt{K}$ and
$\Delta\sigma_2(B)\sqrt{K}$, respectively.
The dashed lines are provided as a guide to the eye.}\label{f6}
\end{figure}

The temperature dependences of $\tau_\phi$ and $\tau_s$ for two gate
voltages are depicted in Fig.~\ref{f6}. One can see that the $T$
dependences of $\tau_\phi$ obtained for each of the arms coincide very
closely. They are well described by the $1/T$ law that is consistent
with theoretical prediction for the case when the phase relaxation is
determined by inelasticity of electron-electron (\emph{e-e})
interaction.\cite{AA85} The spin relaxation time  is practically
independent of the temperature as it should be for the degenerate gas
of carriers. Such dependences of the fitting parameters $\tau_\phi$ and
$\tau_s$ support using of the HLN expression for description of
interference induced magnetoresistance. 	

\begin{figure}
\includegraphics[width=0.8\linewidth,clip=true]{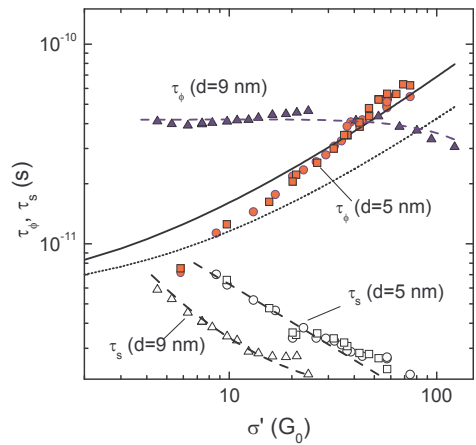}
\caption{The conductivity dependence of $\tau_\phi$ and $\tau_s$
 for HgTe quantum well with inverted ($d=9$~nm) and normal ($d=5$~nm) energy spectrum obtained
 in Ref.~\onlinecite{Minkov12} and this paper, respectively. The solid and dotted lines are calculated
 according to Ref.~\onlinecite{Zala01} with $F_0^\sigma=0$ and $-0.5$, respectively. The dashed lines are
 provided as a quid to the eye.}\label{f7}
\end{figure}

Let us now consider the main result of the paper. It is the
conductivity dependence of the phase relaxation time shown in
Fig.~\ref{f7}, where $\sigma'=\sqrt{\sigma_1\sigma_2}$. It is evident
that $\tau_\phi$ found experimentally increases with the increasing
conductivity. Such the behavior is  analogous to that observed in
quantum wells with ordinary spectrum (see, e.g.,
Ref.~\onlinecite{Min04-2}, where the data for
GaAs/In$_{0.2}$Ga$_{0.8}$As/GaAs quantum well are presented). Our data
are in satisfactory agreement with the theoretical results obtained in
Ref.~\onlinecite{Zala01} for the usual 2D systems for the case when
inelasticity of \emph{e-e} interaction is the main mechanism of phase
relaxation. This is clearly seen from Fig.~\ref{f7}, where the curves
represent the calculation results for two values of the parameter
\emph{e-e} interaction $F_0^\sigma$: $F_0^\sigma=0$ and $-0.5$. Note,
the fact that  most of the experimental points fall between the
theoretical curves cannot be regarded as a method of determining
$F_0^\sigma$~vs~$\sigma$ dependence in these systems.

The growing $\tau_\phi$~vs~$\sigma$ dependence observed in  the present
paper for the narrow quantum well, $d=5$~nm, with normal energy
spectrum differs drastically from that obtained in
Ref.~\onlinecite{Minkov12} by the same method for the structures with
the wider quantum well, $d=(9-10)$~nm, with the inverted subband
ordering. The results from Ref.~\onlinecite{Minkov12} are presented in
Fig.~\ref{f7} also. As seen $\tau_\phi$ is practically independent of
the conductivity in the quantum well with inverted spectrum   in
contrast to the data obtained for the wells with normal spectrum.

Thus,  the fitting parameter $\tau_\phi$ identifying  with the phase
relaxation time behaves itself with increasing $\sigma$ in narrow
quantum well HgTe with normal subband ordering in the same way as in
usual 2D systems. When we are dealing with the electrons in the
inverted band, the $\tau_\phi$~vs~$\sigma$ behavior is extraordinary.
As was discussed in Ref.~\onlinecite{Minkov12} one of the possible
reason for the latter case is that the fitting parameters $\tau_\phi$
and $\tau_s$ may not correspond to the true phase and spin relaxation
times, respectively, despite the fact that the standard HLN expression
fits the experimental magnetoconductivity curves rather well. We
presume that this could result from the fact that the HLN expression
does not take into account a multi-component character of the electron
wave functions. Because electrons in narrow ($d<d_c$) and wide
($d>d_c$) HgTe quantum wells belong to the different branches of the
spectrum\cite{Gerchikov90,Bernevig06} with sufficiently different
structures of the wave functions, the effects of electron interference
can in principle be strongly dependent on the width of quantum well,
which controls  the subband ordering.

Strictly speaking, the above allowance for the conductivity anisotropy
is rigorous in the case when the electric current flows along the
principal axes of the conductivity tensor. We show in the supplemental
material that the measurements performed on the L-shaped Hall bar allow
ones to find the principal exes of the conductivity tensor and, thus,
perform the analysis of the magnetoconductivity for the arbitrary
orientation of the bar and conductivity tensor. It is shown that
misorientation between the conductivity tensor and the Hall bar is
small enough so that the taking it into account is not needed in our
concrete case.

\section{Conclusion}

Investigating the interference induced magnetoconductivity in narrow
HgTe single quantum well we have shown that the phase relaxation time
found from the fit of the low-field magnetoconductivity for the quantum
well with the normal energy spectrum increases  with the conductivity
increase analogously to that observed in ordinary single quantum wells.
Such the behavior is in agreement with that predicted theoretically for
the case when inelasticity of \emph{e-e} interaction is the main
mechanism of the phase relaxation time. At the same time, it differs
markedly from  the behavior of $\tau_\phi$ obtained in wider HgTe
quantum well with inverted energy spectrum,\cite{Minkov12} where
$\tau_\phi$ is nearly constant over the wide conductivity range. It is
proposed that different structure of the multicomponent electron wave
functions for $d<d_c$ and $d>d_c$  could be responsible for such a
discrepancy.

\section*{Acknowledgments}
We are grateful to I.V. Gornyi for numerous illuminating discussions.
This work has been supported in part by the RFBR (Grant Nos.
11-02-12126, 12-02-00098, and 13-02-00322).

\section*{Supplemental Material}

\begin{figure}
\includegraphics[width=0.9\linewidth,clip=true]{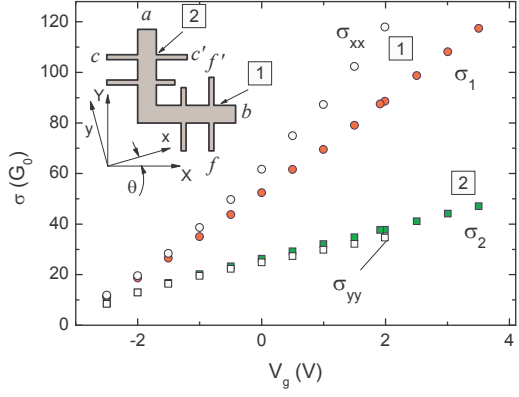}
\caption{The gate voltage dependence of $\sigma_1$ and $\sigma_2$ measured for arms $1$ and $2$
(solid symbols) as compared with that of $\sigma_{xx}$ and $\sigma_{yy}$,
found as described in the text. The inset illustrates the orientation
of the principal axes $x$ and $y$ of the conductivity tensor in the coordinate system of the sample. }\label{fs}
\end{figure}

Analysis in the paper is performed under assumption that the principal
axes $x$ and $y$ of the conductivity tensor coincide with the axes $X$
and $Y$ of the  coordinate system connected with sample as shown in
Fig.~\ref{f2}(a) and Fig.~\ref{f3}. If it is not the case, i.e., the
angle $\theta$ between the $X$ and $x$ axes  is nonzero (see the inset
in Fig.~\ref{fs}), the  $\sigma_{xx}$ and $\sigma_{yy}$ components as
well as the $\theta$ angle can be found from resolving the following
system of equations:
\begin{eqnarray}
\sigma_1^{-1}&=&\sigma^{-1}_{xx}(\cos{\theta})^2+\sigma^{-1}_{yy}(\sin{\theta})^2 \nonumber \\
\sigma_2^{-1}&=&\sigma^{-1}_{xx}(\sin{\theta})^2+\sigma^{-1}_{yy}(\cos{\theta})^2 \nonumber \\
V_{cc'}&=&\frac{1}{2} I_{ab}(\sigma^{-1}_{xx}-\sigma^{-1}_{yy})\sin{2\theta}\nonumber \\
V_{f'f}&=&V_{cc'}, \label{eqs}
\end{eqnarray}
where  the  current $I_{ab}$ flowing through the probes $a$ and $b$,
and the quantities $\sigma_1$, $\sigma_2$, and $V_{cc'}$ (or $V_{f'f}$)
are measured experimentally. Doing so we have obtained $\theta=(22\pm
5)^\circ$, and $\sigma_{xx}$, $\sigma_{yy}$ plotted against the gate
voltage in Fig.~\ref{fs}. One can see that the data obtained are
consistent  with the results obtained in the main paper;
$\sigma_{xx}>\sigma_1$ while $\sigma_{yy}<\sigma_2$ so that
$\sigma_{xx}/\sigma_{yy}>\sigma_{1}/\sigma_{2}$.

As for the interference induced magnetoconductivity measured on the
L-shaped Hall, it is easy to show that the following expression is
valid for $\Delta\sigma_{1,2}(B)$:
\begin{eqnarray}
\Delta\sigma_1(B)&=&\sqrt{\frac{\sigma_1}{\sigma_2}}\,G_0\, {\cal H}\left(\frac{B}{B'_{tr}},\frac{\tau'}{\tau_\phi},\frac{\tau'}{\tau_s}\right)F(K,\theta),\nonumber\\
\Delta\sigma_2(B)&=&\sqrt{\frac{\sigma_2}{\sigma_1}}\,G_0\, {\cal H}\left(\frac{B}{B'_{tr}},\frac{\tau'}{\tau_\phi},\frac{\tau'}{\tau_s}\right)F(K,\theta),
 \label{eqs10}
\end{eqnarray}
where $K=\sigma_{xx}/\sigma_{yy}$ and
\begin{equation}
 F(K,\theta)=\sqrt{\frac{K}{\left(\sin^2\theta+K\cos^2\theta\right)\left(K\sin^2\theta+\cos^2\theta\right)}}.
\label{eqs15}
\end{equation}
Thus, the experimental dependences
$\Delta\sigma_1(B)\sqrt{\sigma_2/\sigma_1}$ and
$\Delta\sigma_2(B)\sqrt{\sigma_1/\sigma_2}$ measured for arbitrary
orientation of the L-shaped Hall  should be described by the same
function ${\cal H}(b,x,y)$ multiplied by the factor $F(K,\theta)$ given
by Eq.~(\ref{eqs15}).

If one uses  $\theta=22^\circ$ and maximal under our conditions value
of $K$, $K=4$, one obtains $F(K,\theta)\simeq 0.9$. Because this value
is close to one, it is obviously that ignoring $F(K,\theta)$ in the
main text does not affect the results of the paper. It is justified by
the direct inclusion of the factor $F(K,\theta)$ into the fitting
procedure. Taking the factor $F(K,\theta)$ into account results in the
maximal increase of $\tau_\phi$ by approximately $2$~\% and decrease of
$\tau_s$ by $20$~\% or less over the entire temperature and
conductivity ranges used in the paper.

\end{document}